# LADIR REPORT 2010/11

**Les cendres végétales, matières premières des verres et des émaux : un exemple, les cendres de végétaux utilisées par Fr. D de Montmolin.**

Ph. Colomban*, A. Tournié
Laboratoire de Dynamique, Interaction et Réactivité - UMR7075
CNRS
Université Pierre-et-Marie-Curie (UPMC)
2 rue Henry-Dunant 94320 Thiais, France

*corresponding author
tel +33 1 4978 1105
Fax +33 1 4978 1118
Philippe.colomban@glvt-cnrs.fr

Résumé:

De part le caractère pulvérulent et leur haute teneur en alcalis les cendres végétales constituent des matières premières pouvant être utilisées comme fondant de compositions silicatées (verres, émaux, céramiques). Si leur utilisation est décrite dans la littérature sur l'histoire des verres depuis des millénaires, les travaux sur l'analyse de leur composition restent limités. Nous discutons ici des compositions de cendres d'arbres et d'arbustes (acacia, aubépine, chêne, chêne vert, olivier, orme, peuplier, pommier, sarment (vigne)), de plantes (carex, quenouille, fougère), de céréales (blés, maïs, riz), déchets de battage et foin, principalement récoltés en Mâconnais, dans les environs de Taizé par le potier Fr. D. de Montmolin. Les apports en fondants alcalins ($Na_2O$, $K_2O$), alcalino-terreux (CaO, MgO) et en silice sont discutés à la lumière des données de la littérature sur l'histoire des techniques verrières et céramiques.
Mots clés : cendres, bois, plantes, céréales, composition, verre, émaux, Raman

Abstract
The powdery nature and high alkali content of vegetable ashes make them ideal raw materials to be used as modifiers of silicate compositions (glasses, enamels and ceramics). Their utilisation since ancient times is described in the literature of the history of glasses, but studies on the analyses of their composition are still limited. We discuss here the compositions of tree and shrub ashes (wattle, hawthorn, oak, green oak, olive wood, elm, poplar, apple tree, vine shoot), of plants (carex, fern, dogwood), of cereals (wheat, maize, rice), threshing waste and hay, mainly harvested in Maconnais, near Taizé (Saône-et-Loire, France), by the potter Brother D. de Montmolin. The contributions in alkali modifiers ($Na_2O$, $K_2O$), alkaline-earth (CaO, MgO) and in silica are discussed in view of the data gathered from the literature of the history of techniques used in the production of ceramics, enamels and glasses. The huge variation in composition is usually attributed to recycling and is questioned by the very broad range of compositions that we obtained in the analyses of the ashes.

INTRODUCTION

Outre la maîtrise du feu (conduite de la température et des conditions d'oxydo-réduction) les verriers et potiers se doivent de contrôler et de favoriser la fusion de certaines des matières premières pour initier la fusion totale (verre, émaux) ou partielle (poterie) d'un mélange silicaté. La solution la plus simple est de jouer sur l'hétérogénéité des compositions des matières premières – l'atmosphère de cuisson - de favoriser la formation de l'eutectique ayant la température la plus basse dans le diagramme de phase. C'est ainsi que les premiers potiers en recherchant des cuissons très réductrice pouvaient obtenir des tessons bien densifier en dessous de 800°C grâce aux eutectiques entre FeO et les alcalins/alcalinoterreux. Des températures supérieures 870 à 1050°C [Levin EM et al., 1969, ibid 1975] sont requises en atmosphère oxydante. L'homogénéisation des verres et émaux nécessite de dépasser les températures du liquidus. Dans tous les cas, il est nécessaire de disperser au mieux les grains de matières premières, en maximisant les contacts de(s) eutectique(s) avec les grains les plus réfractaires. La disponibilité des matières premières en particules fines est donc capital : d'où l'usage d'argiles, de calcaire traité thermiquement (coquilles, craies et marnes), de sable fin, de silex et cristal de roche étonnés[1], … mais aussi de tartre calciné (apport de potasse à partir généralement de la lie de vin), de sel marin et de cendres végétales ou animales (os)! De nos jours, les cendres résultant de certaines industries (condensat de fumée sont utilisées comme ajout dans les ciment ou dans certaines productions céramiques [Rawlings *et al.,* 2006 ; Merino *et al.,* 2005]. Les cendres volcaniques sont aussi été utilisées depuis des siècles [Ford & Rose, 1995].

Comme le résume la Fig. 1, différents auteurs [Sheridan et al., 2005; Artioli et al., 2008] rapportent que les premiers verres de l'Age du Bronze européen étaient préparés à partir de cendres de bois et de plantes. Le Monde méditerranéen et le Bas Moyen-Age européen [Brill, 1999 ; Tite & Shortland 2003] privilégièrent l'usage du Natron, une évaporite disponible dans les chotts des régions désertiques du Proche-Orient comme le *Wadi Natrum* [Shortland, 2004; Shortland *et al.*, 2006], matière déjà utilisée par les Dynasties Egyptiennes pour de nombreuses applications : sa composition proche de $Na_2CO_3, HNaCO_3, H_2O$ en fait une source assez pure de sodium. Avec la fin du Monde Romain et de la *Pax Romana*, la disponibilité du Natron devient très limité et de nouvelles sources locales d'alcali furent privilégiées [Cox *et al.,* 1979; Wedepohl, 1997; Henderson, 2002], en particulier l'usage de cendres de bois et de fougères comme rapporté dans l'ouvrage du moine Théophile [Cannella, 2006]. Cependant au Moyen-Orient des plantes halophytiques de la famille des Chenopodiacae (salona ou salicorne) restent utilisées [Henderson, 2002]. Le marqueur utilisé pour déceler l'usage de cendres végétales est le niveau de potassium (typiquement >1,5% dans un verre sodique); les teneurs en phosphore (bois), magnésium (plantes halophiles) voire en calcium sont aussi des indices pertinents. Aussi, comme le résume la Fig. 2, la représentation des compositions relatives dans les diagrammes des fondants $Na_2O$-$CaO$-$K_2O$+$MgO$ ou des majeurs $SiO_2$-$CaO$-$K_2O$ sont utilisables pour différentier le type de matière première utilisée pour l'élaboration du verre [Brill, 1999; Turner, 1956; Tite & Shortland, 2003]. Ces conclusions largement admises dans la littérature donnent à notre avis une vue schématique et les compositions n'appartenant pas vraiment aux domaines de référence ont été explicitées par des réutilisations de verres d'origine diverses. Si cette proposition est bien étayée par la découverte d'un commerce de calcin coloré par des éléments rares (verres bleus au cobalt par exemple, [Gratuze *et al*., 1992; Gratuze *et al.* 1996]), et ce dès les époques romaines voire phénicienne [Foy, 2003; Colomban *et al.,* 2003], l'étendue de ce recyclage est difficile à

---

[1] Porté à haute temperature et jeté dans l'eau froide le galet se fragmente ou même devient pulvérulent

évaluer. La localisation des compositions rapportées dans la littérature dans les diagrammes des principaux fondants ($Na_2O$-$CaO$-($K_2O$+$MgO$), **Fig. 2**) illustre que si la variété des compositions est grande, des familles bien distinctes peuvent être distinguées, aussi bien à partir des compositions élémentaires que des paramètres Raman caractéristique de l'organisation nano-structurale de la charpente silicatée [Colomban, 2009].

Les cendres ont des compositions très variées, influencées par la nature de l'espèce végétale sélectionnée, la partie prélevée (feuillage, branche, tronc, …), la saison et le sol ou les végétaux ont poussé [ Brill R.H., 1999 ; De Monmollin D., 1997 ; Oppenheim A. L. *et al.,* 1970 ; Turner W. E. S., 1956, Lambercy E., 1993]. C'est pourquoi la possibilité d'analyser un stock important de cendres végétales préparées pour la réalisation d'émaux par Frère Daniel de Montmollin et Frère Lutz Krainhoefner offre une bonne base de discussion pour appréhender les compositions vitreuses pouvant être préparées à partir de cendres végétales.

COMPOSITIONS ELEMENTAIRES

Les cendres végétales, tamisées, (lavées dans un grand excès d'eau pour les cendres acides ou siliceuses, c'est à dire issues de graminées) puis séchées à l'air ont été stockées en sachets de plastique fermés. Les cendres de bois et de végétaux herbacés comme les crucifères et les papilionacés peuvent être stockées sans lavage. La couleur des différents échantillons varie du gris clair au noir (Tableau 1) selon la teneur en carbone. Ces poudres étant fortement hydroscopique des mesures par thermobalance ont été effectuées et les pertes au feu à 150°C effectuées afin de pouvoir calculer les compositions élémentaires sur produit "sec".

Les teneurs élémentaires en S, Si, Al, Fe, Ti, Ca, Mg, K, Na, P et C ont été déterminées par le Service Central d'Analyse du CNRS (Vernaison, France) par les méthodes standard sur produits séchés à 150°C, ce qui garantie une teneur en eau résiduelle inférieure à quelques %. Elles ont été converties en équivalent oxyde (**Tableau 1**). Les teneurs en carbones ont été mesurées sur une sélection de cendres ayants des niveaux de coloration progressif : les teneurs en carbone vont de quelques % pour les échantillons gris jusqu'à près de 10% sur les cendres "noires".

L'analyse des teneurs en silice confirme que le riz est une source très pure en silice (près de 94% - c'est d'ailleurs la seule cendre végétale qui est à notre connaissance utilisée industriellement [Bondioli F. *et al.,* 2007]) mais il en est de même pour les épis de maïs (50 à 75%), pour le blé (50-65%, le grain est très pauvre en silice !) ou le Carex, principal constituant des tourbières (67%), la cornouille (~70%) et de 46 a 52% pour le peuplier, le foin et la fougère. Le bois d'acacia (~7%) mais aussi certains foins (5-6%) et le peuplier (~5%) apportent significativement de l'alumine. Certaines vignes peuvent donner des cendres très riches en alumine, près de 30% [Lambercy E. 1993]. La vigne (sarments), le carex, le foin et le blé ainsi que l'acacia apporte de l'oxyde de fer (2,5-3%). L'apport en oxyde de calcium est dominant dans le bois : 85% pour le chêne vert, 70% pour l'olivier, 60% pour les autres arbres, les teneurs faibles étant observées pour la vigne (45%), l'aubépine et le peuplier (~30%), le foin et fougère (~20%), l'acacia, certaines plantes (carex, cornouille) et les céréales (<15%). La cendre de bois apporte donc beaucoup plus de calcium que la cendre d'os (~50-55%). Les teneurs en potassium sont plus resserrées, entre 10 et 15% pour la plupart des cendres à l'exception de celles d'acacia et d'aubépine (25 à 30%), du trèfle rouge en fleur (35%, Lambercy E., 1993), de certains blé, chêne et fougère (15-20%) et du chêne vert, maïs, carex, olivier, orme et riz (<15%). Le record d'apport de potassium semble obtenu par les cendres de tubercule de pomme de terre avec près de 60%, la balance étant obtenue principalement avec $P_2O_5$ (16%) et MgO (5%) [Lambercy E., 1993]. Autrement, les teneurs en oxyde $P_2O_5$ sont maximale pour les déchets de battage (~20%) et atteignent 5-10% pour la plupart des cendres d'arbres et d'arbustes (acacia, pommier, orme, aubépine, sarments), quelques plantes (carex, cornouille), foin, de certains blés et le maïs. La dispersion des teneurs en MgO est assez réduite (2-5%) à l'exception de l'acacia (9%), la fougère (8%) et l'aubépine, certains blés et les déchets de battage (5-6%).

La variabilité des teneurs en phosphore montre qu'un faible niveau de cet élément n'est en rien une preuve d'un non-usage de cendres végétales. Les cendres végétales semblent être une source importante en calcium. Il est clair que l'apport en sodium est négligeable avec la végétation continentale.

IDENTIFICATION DES MATIERES PREMIERES

La Fig. 3 résume les domaines de composition relative en fondant et en fondants/silice selon les cendres végétales. De façon simplifiée on voit que l'usage de bois favorise les compositions riches en calcium tandis que l'usage d'herbes ou de céréales conduit à des compositions plus riches en potassium. Cependant les teneurs plus importantes en silice de ces dernières limitent l'effet fondant de l'excès en potassium.

Les niveaux très bas de sodium dans les cendres végétales rendent impossible de préparer sans apport supplémentaire de sodium les verres des compositions de la Fig. 2. Il s'ensuit que les verres préparés à partir de cendres "continentales" utilisent d'autres matières premières : il peut s'agir de sable de rivière contenant des feldspaths sodiques (Albite, Sanidine par exemple) et/ou d'apport de sel marin NaCl ou de sulfate de sodium. L'analyse des textes anciens apporte des réponses partielles. L'ouvrage de A.-F. Cannella [Cannella, 2006] qui analyse outre le manuscrit de Jean d'Outremeuse, l'ensemble des textes de recettes verrières note la mention régulière de l'utilisation de sel ou poudre d'alcali (p 126-129). L'utilisation de sulfates semble difficile car l'incorporation de sulfate fondu dans le verre fondu est difficile. Cependant la nature réelle des réactifs restent discutables, en particulier ceux dénommés comme "tartre calciné ou de soude" (sic). Jean d'Outremeuse indique l'usage de "sel alcali bien blanc" (p 132, [Cannella, 2006]) obtenu à partir de plantes. Il semble donc établi que des cendres de plantes côtières, riche en sodium aient été disponibles et donc en complément des cendres "continentales" permettent d'atteindre l'ensemble des compositions notées "plantes continentales" de la Fig. 2a, domaine notoirement plus large que celui de la Fig 3a obtenu avec les cendres du Tableau 1. Cannella rapporte que selon M. Verita [Verita, 1991c p133] la soude est importée du Moyen-Orient (Syrie-Egypte) à Venise (*alume catino, cenere di Levante, cenere di Soria*) et consiste en cendres de *salsola soda* ou kali.

La Fig. 4 présente les localisations des compositions de vitraux de cathédrales françaises et allemandes [Sterpenich, 1998] ainsi que les liquidus et leur températures à partir du diagramme de phase [Morey *et al*., 1930]. Ces domaines de composition sont cohérents avec d'autres travaux [Schalm *et al.,* 2004]. Les compositions des vitraux allemands du XIVe siècle riches en calcium et ceux français/allemand des VII-XIVe siècles riches en potassium , sont en très bon accord avec une préparation à partir de cendres "continentales" tandis qu'il est évident que les vitraux de Rouen des VII-IXe siècles riches en sodium ont été préparés à base de soude (sel ou cendres de plantes côtières ou de lac salés importées).

ANALYSE NON-DESTRUCTIVE DES VERRES

L'identification des types de composition est réalisée à partir de prélèvements ce qui n'est pas possible pour tous les échantillons. Une alternative efficace est maintenant possible avec la spectrométrie Raman [Colomban, 2008], technique pouvant être réalisée sur site, par exemple sur des vitraux en place [Colomban & Tournié, 2007] comme sur des objets ne pouvant sortir de leur lieu de conservation [Ricciardi *et al.* 2009a; ibid 2009b; Kirmizi *et al.,* 2010a, ibid, 2010b]. Cette identification est obtenue à partir de l'analyse de la nanostructure du réseau silicaté [Si-O-Si-O-Si]n qui dépend principalement de la composition, en particulier des teneurs en fondants (Na, K, Ca) qui détermine le degré de polarisation du réseau silicaté [Colomban, 2008; Colomban *et al.* 2006] et accessoirement de l'histoire thermique. Le spectre Raman est aussi fonction du degré de covalence du réseau de tétraèdre et donc du rapport Silice/alumine par exemple. Il s'en suit que si les compositions de verres ou d'émaux peuvent être différentier par leur taux de fondant (et leur nature) ou indice de polymérisation [Colomban Ph., 2003a ; b] et par les rapports silice/alumine, la spectroscopie Raman sera efficace [Colomban *et al.,* 2006]. La Fig. 5 présente les classifications faites à partir des données de la littérature [Colomban *et al.*, 2006 ; Lagabrielle S. et Velde B., 2003 ; Schalm O., *et al.*, 2004 ; Sterpenich J., 1998 ; Tournié A. *et al*, 2006] pour les vitraux d'édifice religieux de France, Belgique, Pays-Bas et d'Allemagne. Ces diagrammes démontrent qu'aux époques de renouveau technologique (XIVe siècle en Allemagne, XIXe siècle) les compositions sont très variables et donc la spectroscopie Raman différentiera très bien les types de vitraux. Mais cela est aussi possible pour les autres productions car les taux de fondant et/ou le degré de covalence change [Colomban et 2006a; ibid 2006b].

CONCLUSION

La détermination expérimentale des compositions de différentes cendres de plantes "continentales" montre que la variabilité des compositions est très large. Les teneurs très faibles en sodium confirment que la préparation de verres sodiques et même des verres mixtes sodiques-potassiques nécessite le recours à des matières premières riches en sodium (cendres de plantes maritimes importées, natron ou sel marin). Les différences de teneurs en alcalins/alcalino-terreux provenant de l'usage de matières premières différentes sont à l'origine de l'efficacité de l'analyse Raman, technique non-destructive pouvant être réalisée sur site avec des dispositifs mobiles.

REMERCIEMENTS

| | Echantillon | | Référence | Résidu sec / couleur | C | SO$_4$ | SiO$_2$ | Al$_2$O$_3$ | Fe$_2$O$_3$ | TiO$_2$ | CaO | MgO | K$_2$O | Na$_2$O | P$_2$O$_5$ | Total avec C | total sans C |
|---|---|---|---|---|---|---|---|---|---|---|---|---|---|---|---|---|---|
| arbre | Acacia | 1 | ACAC | Beige | 8.58 | 4.46 | 22.31 | 7.25 | 2.45 | 0.47 | 16.53 | 9.25 | 25.34 | 1.02 | 10.92 | 108.58 | 100 |
| | | 2 | SCA 05007883 | | 8.54 | 5.29 | 23.85 | 7.39 | 2.49 | 0.47 | 15.63 | 9.13 | 24.94 | 0.6 | 10.22 | 108.55 | 100.01 |
| | Aubépine | 3 | AUBEP | Gris clair | | 3.55 | 22.34 | 1.71 | 0.76 | 0.11 | 31.39 | 5.33 | 27.45 | 0.89 | 6.47 | 100 | 100 |
| | | 4 | SCA 05007883 | | | 0 | 22.12 | 1.81 | 0.84 | 0.12 | 32.71 | 5.69 | 29 | 0.94 | 6.76 | 99.99 | 99.99 |
| céréales | Blé 04 | 5 | BLE04 | Noir | | 0.46 | 65.15 | 1.01 | 0.91 | 0.06 | 8.69 | 3.34 | 16.67 | 0.21 | 3.5 | 100 | 100 |
| | Blé 76 | 6 | BLE76 | Gris clair | | 0 | 51.41 | 2.89 | 2.28 | 1.29 | 11.19 | 5.8 | 9.93 | 1.32 | 13.89 | 100 | 100 |
| plante | Carex | 7 | CAREX | Gris foncé | 3.89 | 0 | 67.42 | 2.62 | 2.31 | 0.08 | 9.48 | 3.67 | 6.77 | 0 | 7.64 | 103.88 | 99.99 |
| arbre | Chêne | 8 | CHENE | Beige marron | | 5.07 | 13.42 | 1.52 | 1.43 | 0.11 | 54.69 | 2.07 | 18.47 | 0.35 | 2.86 | 99.99 | 99.99 |
| | Chêne vert | 9 | CHEVT | 61.50 Beige | | 0 | 2.44 | 1.32 | 0.34 | 0.05 | 84.5 | 1.83 | 7.52 | 0.09 | 1.92 | 100.01 | 100.01 |
| | | 10 | SCA 05007883 | | | 0 | 2.87 | 1.27 | 0.3 | 0.05 | 84.33 | 1.81 | 7.47 | 0.12 | 1.78 | 100 | 100 |
| déchet | Déchets de battage | 11 | DECBA | Gris | | 3.7 | 37.56 | 2.59 | 1.63 | 0.1 | 16.41 | 5.54 | 10.43 | 0.34 | 21.71 | 100.01 | 100.01 |
| | | 12 | SCA 05007883 | | | 0 | 39.49 | 2.74 | 1.63 | 0.1 | 17.23 | 5.69 | 10.82 | 0.35 | 21.95 | 100 | 100 |
| foin | Foin 70 | 13 | FOIN70 | Gris | | 0 | 48.61 | 4.91 | 2 | 0.24 | 20.68 | 4.24 | 12.6 | 0.85 | 5.87 | 100 | 100 |
| | Foin 78 | 14 | FOIN78 | Gris | | 0 | 52.06 | 6.3 | 2.98 | 0.31 | 17.12 | 3.52 | 10.78 | 1.05 | 5.89 | 100.01 | 100.01 |
| fougère | Fougère | 15 | FOUGR | Gris | | 0 | 48.87 | 1.29 | 1.63 | 0.1 | 18.71 | 8.21 | 16.68 | 0.92 | 3.6 | 100.01 | 100.01 |
| céréales | Maïs 03-CH | 16 | MA3CH | Gris foncé | 7.86 | 0 | 47.72 | 3.01 | 0.52 | 0.34 | 40.43 | 1.91 | 4.42 | 0.31 | 1.33 | 107.85 | 99.99 |
| | Maïs 03-StR | 17 | MA3SR | Gris foncé | | 0 | 53.18 | 4 | 1.49 | 0.23 | 15.07 | 4.11 | 14.14 | 0.31 | 7.48 | 100.01 | 100.01 |
| | | 18 | SCA 05007883 | | | 0 | 52.03 | 4.17 | 1.56 | 0.23 | 15.14 | 4.28 | 14.5 | 0.4 | 7.69 | 100 | 100 |
| | Maïs 73 | 19 | MA73 | 91.00 Gris | | 0 | 76.98 | 3.33 | 0.77 | 0.14 | 6.44 | 2.35 | 6.83 | 0.3 | 2.87 | 100.01 | 100.01 |
| arbre | Olivier | 20 | OLIV | Beige | | 1.63 | 6.21 | 2.95 | 0.7 | 0.14 | 70.62 | 4.7 | 9.32 | 0.58 | 3.16 | 100.01 | 100.01 |
| | Orme | 21 | ORME | Gris clair | | 2.91 | 16.25 | 1.21 | 0.71 | 0.05 | 58.89 | 4.89 | 8.89 | 0.4 | 5.81 | 100.01 | 100.01 |
| | Peuplier | 22 | PEUPL | Gris clair | | 5.85 | 46.01 | 4.76 | 1.32 | 0.18 | 27.3 | 1.9 | 10.15 | 0.58 | 1.95 | 100 | 100 |
| | Pommier | 23 | POM | Beige | | 0.57 | 12.2 | 1.81 | 1.13 | 0.08 | 62.57 | 4.6 | 11.92 | 0.11 | 5 | 99.99 | 99.99 |
| | | 24 | SCA 05007883 | | | 0 | 11.21 | 2.17 | 1.27 | 0.08 | 63.54 | 4.6 | 11.85 | 0.11 | 5.17 | 100 | 100 |
| plante | Quenouille 71 | 25 | QUE71 | 6.00 Gris fon | 9.46 | 0 | 68.76 | 3.23 | 1.04 | 0.14 | 3.46 | 4.96 | 11.34 | 0.2 | 6.86 | 109.45 | 99.99 |
| céréales | Riz | 26 | RIZ | Gris foncé | 0.37 | 0 | 93.81 | 0.22 | 0.2 | 0 | 1.09 | 0.88 | 2.74 | 0.11 | 0.95 | 100.37 | 100 |
| arbre | Sarment | 27 | SARMT | 72.00 Beige marron | | 1.87 | 20 | 4.2 | 3.01 | 0.25 | 46.39 | 4.82 | 14.43 | 0.23 | 4.8 | 100 | 100 |

# Références

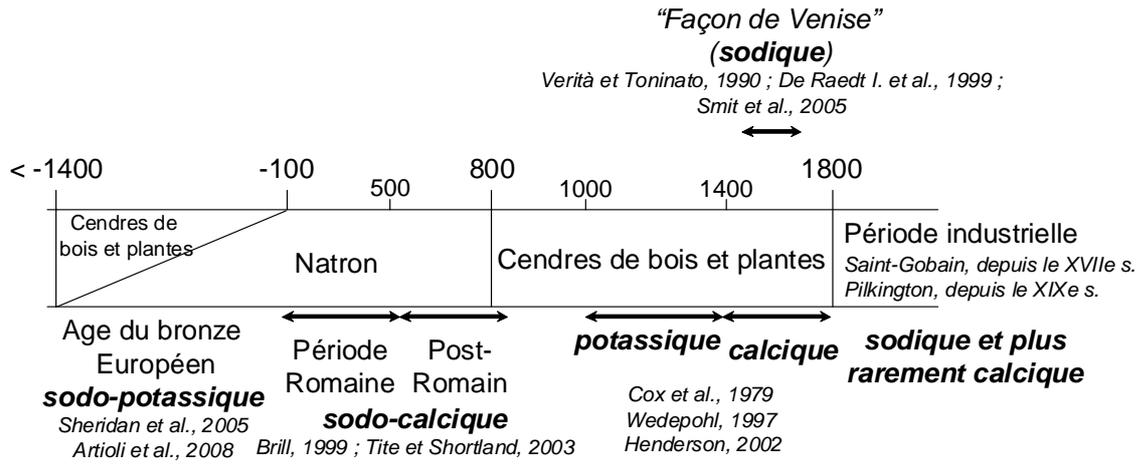

**Fig. 1**: *Principales sources de fondants pour les verres.*

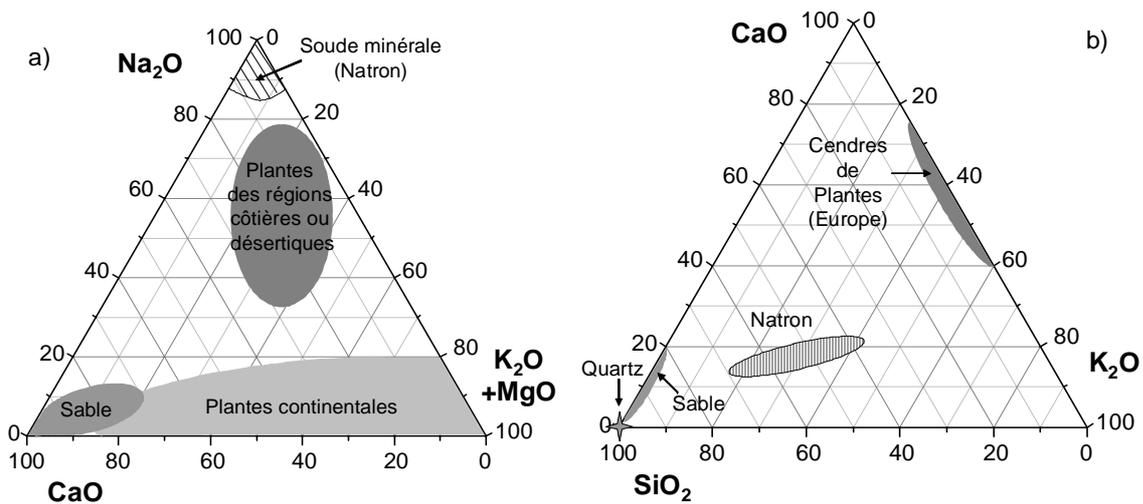

**Fig. 2**: *Diagrammes ternaires des différents types de fondants et formateurs, synthèse à partir des données de Brill R.H., 1999, Turner W.E.S., 1956, Tite M.S. et Shortland A.J., 2003.*

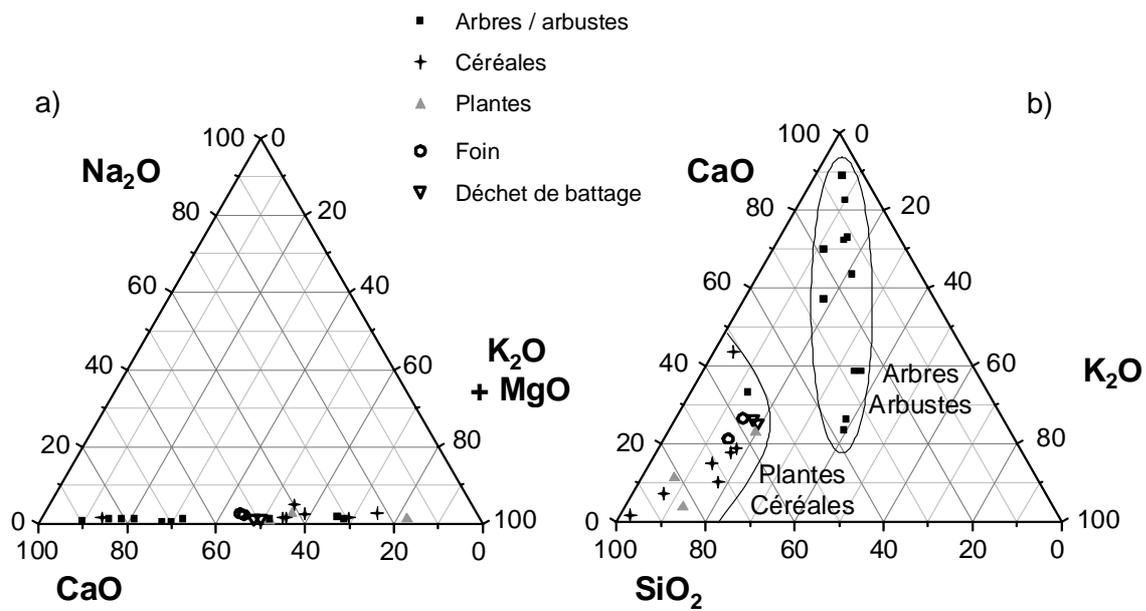

**Fig. 3** : *Diagrammes ternaires de différents types de cendres de végétaux recueillis dans la région de Taizé (Saone-et-Loire) constituée d'arbres/arbustes (acacia, aubépine, chêne, olivier, orme, …), de céréales (blé, maïs, riz) et de plantes (carex, fougère, quenouille).*

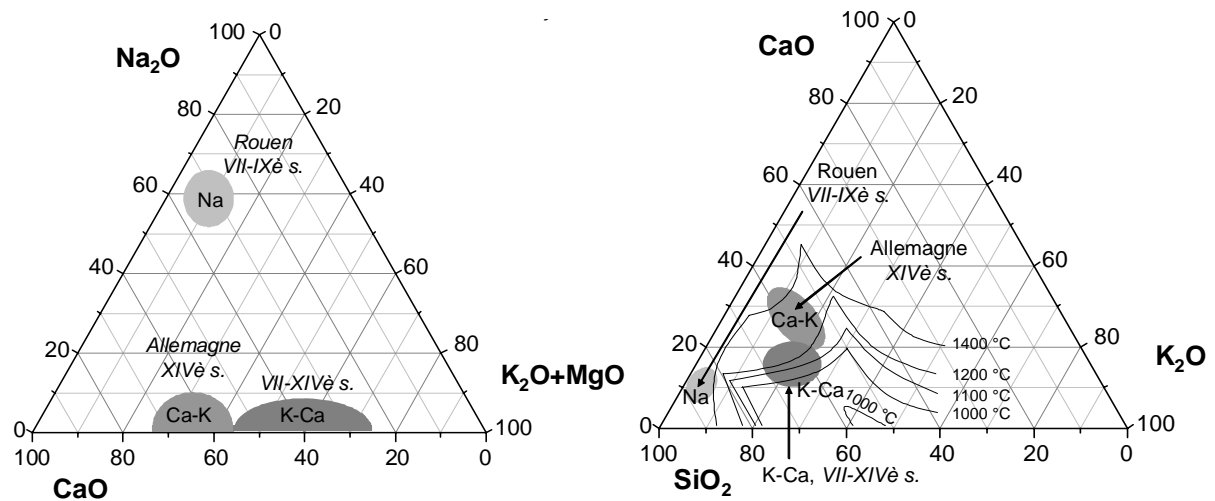

**Fig. 4** : *Comparaison dans les diagrammes ternaires $Na_2O$ – $CaO$ - ($K_2O$+$MgO$) et $CaO$ - $SiO_2$ - $K_2O$ des compositions des vitraux de cathédrales Françaises et Allemandes (VII-XIVe s) étudiés par J. Sterpenich, 1998. Les liquidus sont reportés d'après le diagramme de phase ternaire de Morey et al., 1930.*

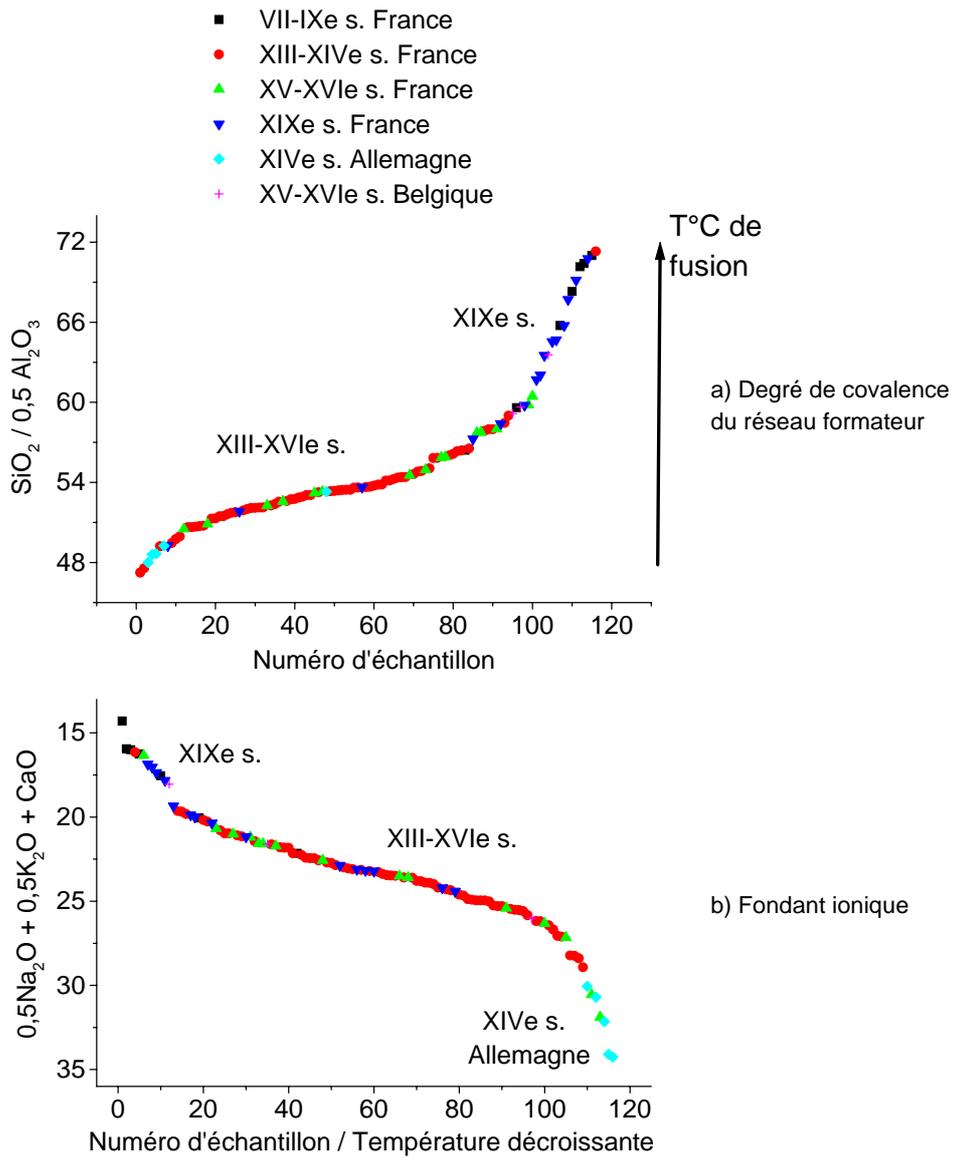

**Fig. 5** : *Hiérarchisation des vitraux à partir **a)** de l'indice de covalence (SiO₂/0,5Al₂O₃) et **b)** de la teneur en fondant ionique (0,5 Na₂O + 0,5 K₂O + CaO), (voir annexe é pour le détail des compositions).*